# Cell Sequence and Mitosis Affect Fibroblast Directional Decision-Making during Chemotaxis in Tissue-Mimicking Microfluidic Mazes

Q. L. Pham,[1] D. Chege,[2] T. Dijamco,[3] J. Brito,[1] E. Stein,[1] N. A. N. Tong,[1] S. Basuray,[1] and R. S. Voronov[1]

[1.] Otto H. York Department of Chemical, Biological and Pharmaceutical Engineering, New Jersey Institute of Technology, Newark, NJ 07102, USA. Email: rvoronov@njit.edu, Fax: +1 973 596 8436, Tel: +1 973 642 4762
[2.] Department of Electrical and Computer Engineering, New Jersey Institute of Technology, Newark, NJ 07102, USA.
[3.] Computer Science Dept., New Jersey Institute of Technology, Newark, NJ 07102, USA

**Abstract**

Directed fibroblast migration is central to highly proliferative processes in regenerative medicine and developmental biology, such as wound healing and embryogenesis. However, the mechanisms by which single fibroblasts affect each other's directional decisions, while chemotaxing in microscopic tissue pores, are not well understood. Therefore, we explored the effects of two types of relevant social interactions on fibroblast PDGF-BB-induced migration in microfluidic tissue-mimicking mazes: cell sequence and mitosis. Surprisingly, it was found that in both cases, the cells display behavior that is contradictory to the chemoattractant gradient established in the maze. In case of the sequence, the cells do not like to take the same path through the maze as their predecessor, when faced with a bifurcation. To the contrary, they tend to alternate - if a leading cell takes the shorter (steeper gradient) path, the cell following it chooses the longer (weaker gradient) path, and vice versa. Additionally, we found that when a mother cell divides, its two daughters go in opposite directions (even if it means migrating against the chemoattractant gradient and overcoming on-going cell traffic). Therefore, it is apparent that fibroblasts modify each other's directional decisions in a manner that is counter-intuitive to what is expected from classical chemotaxis theory. Consequently, accounting for these effects could lead to a better understanding of tissue generation in vivo, and result in more advanced engineered tissue products in vitro.

## Introduction

Directional decision-making during cell migration is important for regenerative medicine (e.g. tissue engineering, wound healing)[1] and developmental biology,[2,3] since tissue development depends on how the cells distribute themselves within the complex pores of the extra-cellular matrix (ECM).[3-5] However, current single cell studies are not representative of the scenarios in which multiple cells enter the pores simultaneously and influence each other's decisions. At the same time, "collective" migration investigations mainly focus on cells that form stable adhesions between each other during movement (e.g. epithelial cells).[6-8] Yet, fibroblasts, which are more relevant to synthesizing collagen and ECM, tend to migrate as individual cells when squeezing through the microscopic tissue pores.[9] As a result, the processes by which single fibroblasts affect each other's directional decision-making in tissues are not well understood. Given the lack of knowledge, this manuscript set out to investigate two unexplored (see Error! Reference source not found. for an overview of recent studies) aspects of fibroblast migration that are particularly relevant to proliferative environments hold true under chemotaxis and in the confines of tissue-mimicking pores.

To that end, we chose the most appropriate migration platform available (see Error! Reference source not found. for an overview of recent devices used to study cell directional decision-making) – a microfluidic maze, which contains tissue-mimicking structural features that offer multiple directional choices.[10,17] Specifically, the maze consists of a long path, a short path, and dead ends. In order to emulate chemotaxis in tissue and induce the fibroblast migration across the maze, we selected platelet derived growth factor – BB (PDGF-BB), a fibroblasts "mitoattractant" (i.e., a chemoattractant[18-20] and a mitogen[21-24] at the same time).

commonly encountered in regenerative medicine: effects of cell sequence and mitosis on chemotaxis in tissue-like pores.

In case of the former, we hypothesized that the order in which the cells enter the pores will affect their directional choices at bifurcations. Fibroblast migration in tissue is primarily regulated via attraction to chemical substances originating from the hematopoietic and blood system, or from products of the ECM. However, other cell types have been known to self-generate/modify existing chemical gradients when confined in microscopic spaces.[10] Moreover, fibroblasts are known to deposit integrin-containing "migration tracks" that affect the motility of their neighbors.[11,12] Therefore, it is logical to conclude that the fibroblast sequence effects will compete with chemotaxis in tissue pores.

Secondly, since tissue generation is a highly proliferative process, we aimed to explore the effects that cell division has on the fibroblast decision-making during chemotactic migration. Mitosis has been found to affect migration in both fibroblasts[12,13] and other cell types.[14-16] However, these experiments were conducted in the absence of a chemotactic gradient and/or micro-confinement. Therefore, we wanted to see whether the cell division effects on fibroblast migration

To our surprise, we found that in both, the cell sequence and the mitosis cases, the cell-cell interactions cause the fibroblasts to display behavior that is contradictory to what would be classically expected from the chemoattractant gradient established in the maze. Namely, we found that the fibroblasts path choices alternate depending on each predecessor's decision, and that cell division occurring during the chemotaxis yields daughter cells with directional bias distinctive from that of their siblings. Therefore, the presented results carry practical implications for both engineered tissue design, and for understanding the fibroblast biology in their native environments.

Table 1. Summary of previous findings on social interactions effects on fibroblast migration behavior

| Interaction Type | Influence on migration | Migration Platform | References |
| --- | --- | --- | --- |
| Collision | Randomize directionality/pseudopod collapse | Non-chemotactic / 2D flat surface | Vedel et al [25] |
| Chemokine-mediated cell signaling | Pseudopod formation bias | | |
| Transient contact | Promote individual cell migration | Non-chemotactic/ 3D Collagen matrix | Miron-Mendoza et al [13] |
| Adhesion | Promote cell clustering | Non-chemotactic / 3D Fibrin matrix | |
| Division | Cells migrate in the same track after division | | |
| Geometrical constraints & physical crowding | Increased alignment and migration speed | Non-chemotactic / Confined fibronectin strips | Leong et al [26] |
| Contact number | Lower migration speed | Non-chemotactic / 2D flat surface | Abercrombie et al [27] |
| Contact inhibition | Immobilize other cells | | Abercrombie et al [28] |
| Adhesion | Promote cell clustering | Non-chemotactic / 3D collagen matrix | da Rocha-Azevedo et al [29] |
| Division | Divided cells migrate in mirror symmetrical tracks | Non-chemotactic / 2D flat surface | Albrecht-Buehler et al [12] Levinstone et al [30] |
| Collision | Cells reflect at the collision area | | Albrecht-Buehler et al [12] |

Table 2. Summary of previous findings on directional decision-making of various types of single cells

| Cell Type | Migration Platform | Directional Cue | Decision Influence | References |
| --- | --- | --- | --- | --- |
| Epithelial cancer cells | Complex maze | Epidermal growth factor (EGF) | Self-generated gradient | Scherber et al[10] |
| Neutrophil | Simple maze | formyl-methionyl-leucyl-phenylalaine (fMLP) | Steeper chemoattractant gradient | Ambravaneswaran et al[17] |
| Neutrophil | Biased bifurcation | Hydraulic resistance + fMLP | Lower hydraulic resistance | Prentice-Mott et al[31] |
| Breast cancer cells | Biased bifurcation | Directional and dimensional constraints | Path size constraint and path orientation; Cytoskeletal drugs | Mak et al[14] |
| Breast cancer cells | Biased bifurcation | Physical cues + contact guidance | Amount of wall contact and path size constraint; | Paul et al[32] |
| Human prostate cancer | Tapered straight channels | EGF + mechanical constraint | Path size constraint | Rao et al[33] |

## Methodology

### Materials

Polydimethylsiloxane (PDMS) Sylgard 184 was purchased from Dow Corning (Midland, MI). Negative photoresist SU-8 was purchased from Microchem (Newton, MA). Bovine collagen Type I 3mg ml$^{-1}$ solution (PureCol) was purchased from Advanced Biomatrix (San Diego, CA). Recombinant rat platelet-derived growth factor-BB (PDGF-BB) was purchased from R&D Systems (Minneapolis, MN). Culture media was prepared from Minimum Essential Medium (MEM) (Sigma, MO) supplemented with 10% (v/v) fetal bovine serum (FBS) (VWR, Radnor, PA), and 1% (v/v) penicillin-streptomycin (10,000 U mL$^{-1}$) (Thermofisher, Waltham, MA). Basal media was composed of MEM supplemented with 1% (v/v) penicillin-streptomycin. For incubation in 5% $CO_2$ atmosphere, media was buffered by 26 mM sodium bicarbonate (Sigma, MO). $CO_2$-independent media buffered by 20 mM HEPES (Sigma, MO) was used for the microscope stage-top experiment.

### Device description

The microfluidic platform used in this work was a tissue-mimicking maze adapted from a cancer migration study by Scherber et al.[10] The maze is a network of bifurcations consisting of two through paths, one short (1,300 μm) and one long (1,700 μm), and several dead ends. Its channels were chosen to have a rectangular profile 24 μm wide by 17 μm high (see Figure 1A & D), in order to limit the number of cells entering the maze to 1-3 at a time. Each maze had a single entrance and exit, shown at the bottom and top of those figures, respectively. Twenty mazes of identical design were replicated per a single microfluidics device (see Figure 1B & E). The device connected a large cell seeding area, with zero initial concentration of the chemoattractant, to a central reservoir of chemoattractant-rich media, via the mazes. In this manner, a stable chemoattractant gradient was established between the cell compartment and the central reservoir, in order to prescribe direction to the cell migration. Finally, multiple identical devices were placed inside of a PDMS enclosure for high-throughput experimentation (see Figure1C & F). The enclosure also isolated the whole setup from the external environment, in order to prevent contamination.

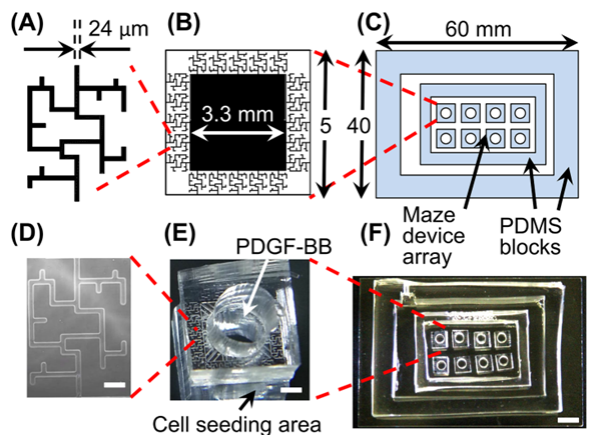

Figure 1. (A, B, C) Schematics showing dimensions of a single maze, a single microfluidic device, and an array of the devices inside of a stage-top custom incubation chamber, respectively. (D) Phase contrast microscopy image of a single maze network. Scale bar is 100 μm. (E) Photograph of a single microfluidic device, consisting of 20 maze networks that share a central chemoattractant reservoir. A circular 3 mm diameter hole was punched in the center of each device in order to deposit the PDGF-BB into the reservoir. Cells were seeded outside of the devices. Scale bar is 1 mm. (F) An array of microfluidic devices placed inside a stage-top custom incubation chamber, consisting of two PDMS blocks bonded to the top of a glass slide. Scale bar is 5 mm.

**Custom incubation chamber for culturing on the microscope stage**

PDMS with a base-to-crosslinking agent ratio of 10:1 was poured into two 100 mm petri dishes to different thicknesses: 8 mm and 10 mm. After heating at 65 °C overnight, the resulting solid PDMS blocks were removed from the petri dishes and cut in rectangular shapes: one smaller and one larger. A 2 mm hole was punched on one side of the larger block for mounting a temperature sensor. The two blocks were then treated with air plasma using a plasma cleaner (PDC-001, Harrick Plasma, Ithaca, NY), and bonded on to a pre-cleaned 50x75x1 mm glass slide (Ted Pella, Redding, CA), one inside the other (see Figure1C & F). The larger and thicker block served as the periphery of the incubation chamber, while the smaller block formed an enclosure for the microfluidic devices array, in order to prevent contamination from the temperature sensor opening. Phosphate buffer saline (PBS) was added to the space between the two PDMS blocks in order to humidify the incubation chamber.

**Device fabrication**

The mold for the device was fabricated using SU-8. First, the microscale pattern was sketched using AutoCAD (Autodesk, Mill Valley, CA) and printed at 16,525 dpi on a transparency (Fineline Imaging, Colorado Springs, CO) to generate a high-resolution photomask. SU-8 was spin coated, exposed to UV light, and developed on a 4-in silicon wafer (University Wafer, Boston, MA) following the manufacturer's protocol to generate 17 μm-high microfluidic channels. Microfluidic devices were fabricated using a single-layer soft lithography method. Typically, PDMS with a base-to-agent ratio of 10:1 was poured over the photo-patterned mold, degassed, and cured at 65 °C overnight. The cast PDMS was carefully peeled off from the master, and a 3 mm biopsy punch was used to create a central hole in each device for PDGF-BB delivery to the chemoattractant reservoir. Individual devices were cut using a thin razor blade, rinsed with 70% isopropyl alcohol and left dried in an oven. Then 6 to 8 devices were treated with air plasma for 30 s at medium radio-frequency power, before they were bonded inside of the culture chamber. The assembly was then heated to 65 °C on a hot plate for 5 min to improve the bonding.

**Surface treatment**

Right after bonding, the device was placed under 200 mTorr strong vacuum for 2 min to remove air inside the micro channels. Immediately after being released from vacuum, about 150 μg ml$^{-1}$ collagen type I solution Purecol was added to completely coat the micro channels and the cell seeding area. The coated device was then sterilized by UV light inside of a biohood for 1 hour, and washed several times with PBS before use.

**Cell preparation**

Mouse embryo NIH/3T3 (ATCC® CRL-1658™) fibroblasts were purchased from ATCC (Manassas, VA). Prior to being transferred to the microfluidic device for the migration experiments, the cells were incubated in the culture media inside of T75 flasks. The flasks were kept at 37 °C and in a humidified atmosphere of 5% $CO_2$ in air. The culture media was changed every two days to ensure normal cell growth. Prior to the migration experiments, the cells were trypsinized from the T75 flasks and loaded into the chip, with a seeding density of about 50,000 cells cm$^{-2}$. The chip was incubated at 37 °C under 5% $CO_2$ for 6 h to allow cell attachment. Then the cells were cultured in serum-starved media (MEM supplemented with 1% penicillin-streptomycin) for 6 hours.

**Cell migration experiments and image acquisition**

At the start of the experiment, cell culture media in the chip was replaced with $CO_2$-independent basal media buffered by

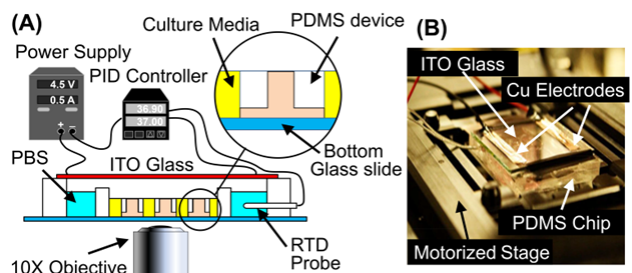

Figure 2. (A) Experiment setup schematic, showing the custom incubation chamber used for culturing on the automated microscope's stage. The setup consisted of a benchtop variable power supply, which provided a current of 0.5 A to an ITO glass that was placed on top of the PDMS chamber. Heat generated by the glass warmed up the media inside of the chamber. Temperature was measured using a RTD sensor coupled to a PID controller, which modulated the power supply in order to maintain a constant temperature of 37 °C. (B) Photograph of the incubation chamber in operation. Electric current was fed to the resistive layer of the ITO glass through a pair of copper electrodes. This led to heat generation underneath of the PDMS chamber. The incubation chamber was placed on top of a XY motorized stage, controlled by the computer, for imaging.

HEPES. 20 μL of basal media supplemented with 25-50 ng mL$^{-1}$ PDGF-BB was then added into the central reservoir of each device. The concentration of PDGF-BB at the exit was chosen to be higher than that typically used to induce migration and mitosis,[22] keeping in mind that it would be lower at the maze's entrance. The experimental setup used a custom incubation system, which is illustrated in Figure 2. A resistance temperature detector (RTD) sensor (Auberin, Alpharetta, GA) was mounted to the mounting hole on the larger PDMS block. The chip was covered by a 15 $\Omega$ cm$^{-2}$ Indium Tin Oxide (ITO)-coated glass slide (Adafruit, NY) and mounted on a motorized microscope stage (Ludl Electronic, Hawthorne, NY). Electrodes on the ITO glass slide were connected to a bench top power supply (MPJA, FL) to supply a heating power of 1.5 W. The RTD sensor was connected to a PID controller (Auberin, Alpharetta, GA) which helped to maintain a constant temperature of 37 ± 0.2 °C. Time-lapse phase-contrast imaging of the fibroblast migration was performed using a fully automated Olympus IX83 microscope fitted with a 10X phase-contrast objective (Olympus, Japan), a CMOS camera (Orca Flash 4.0 V2, Hamamatsu, Japan), and an autofocus module (ZDC, Olympus, Japan). Time-lapse images were automatically captured at 5-15 min intervals for a duration of 20 hr. For each device at each time step, 36 tile images were acquired at different locations, stitched, and stabilized using an in-house Matlab® 2016b code (The MathWorks, Inc., Natick, MA).

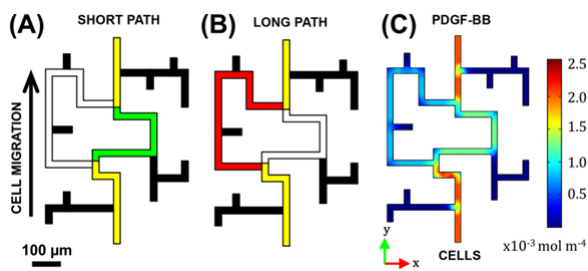

**Figure 3**. Illustration of possible path choices available to the cells in the microfluidic maze. Yellow = common paths; Black = dead ends. (A) Short path labelled in Green; (B) Long path labelled in Red; (C) PDGF-BB gradient magnitude calculated from COMSOL. The maze entrance is at the bottom, and its exit is at the top.

**Data Analysis**

Migratory cells were tracked using the Manual Tracking plug-in for ImageJ software (National Institutes of Health).[34] The directional decision chosen by each individual cell at the long-short bifurcation was determined via manual observation. Quantitative data of cell sequence was generated using an in-house Matlab® 2016b code (The MathWorks, Inc., Natick, MA). Cells that underwent division were not included in the sequential cell migration statistics. Instead, the directional decisions of the mitotic cells were counted separately.

**Concentration gradient simulation**

The PDGF-BB concentration gradient formed between the two ends of the maze was simulated numerically using COMSOL 5.2a Multiphysics (COMSOL, Burlington, MA). Specifically, three dimensional transport of diluted species was modelled inside of 24-µm wide by 17-µm high channels, resembling the real maze geometry (see Figure 3). The chemoattractant reservoir and the cell seeding area were treated as infinite sources and sinks, respectively. Constant concentration boundary conditions, corresponding to the experiment, were set to 50 ng mL$^{-1}$ PDGF-BB at the source (maze exit), and to zero at the sink (maze entrance). A diffusion coefficient of $1.08 \times 10^{-6}$ cm$^2$ s$^{-1}$ was used to represent the PDGF-BB (MW = 25 kDa) in an aqueous solution.[35] Chemoattractant gradient values at every point in the maze were calculated from the vector magnitude of two components, x-axis (horizontal) and y-axis (vertical), both of which increased towards the higher concentration of PDGF-BB (see Figure 3C). The presence of the cells in the maze was not taken into account by the calculation.

## Results

To better understand how single fibroblasts affect each other's directional decision-making under chemotaxis in microscopic tissue pores, we studied their motility in tissue-mimicking microfluidic mazes. The mazes contained two possible through paths and multiple dead-ends. The cells entered the mazes through a common path, and would eventually reach a bifurcation where they had a choice of either a shorter through path (see Figure 3A), or one that was approximately 1.3 times longer (see Figure 3B). Eventually, they exited the maze through another common path (assuming that they did not get stuck in a dead-end).

Since the fibroblasts migrated in response to a concentration gradient of PDGF-BB established inside of the maze, we characterized the relative difference in the chemotactic driving forces between the two paths via a COMSOL simulation (see Figure 3C). The chemoattractant concentration in the maze increased from the zero boundary condition at its entrance (where the cell seeding area was) to 50 ng mL$^{-1}$ at its exit, in order to match the experiment. The resulting gradient, shown in Figure 3C, was about 1.5 times higher in the short path than the long path, mainly due to the length difference between them. Therefore, it would be logical to expect that the cells would be more likely to select the shorter path, in order to ascend the steepest gradient possible.

To test whether the above prediction is valid, the cells were allowed to migrate across the maze in response to two different PDGF-BB concentrations at the exit: 25 and 50 ng mL$^{-1}$. Manual tracking was used to collect statistics on their positions in the maze, throughout the duration of each experiment (see Video S1). Moreover, we also recorded the sequence in which the cells entered the maze and made their directional decisions once inside.

Overall, we found no statistically significant differences between the two PDGF-BB tested concentrations; and the shorter path was indeed preferred over the longer path, for the leading cells that were the first to arrive at the bifurcation (see Figure 4A). The distribution between the short and the long paths was approximately 60-40%, when only these cells were considered. However, surprisingly, if all the cells entering the maze were taken into account (regardless of their order), the directional bias disappeared, and the prediction was no longer valid (see Figure 4B).

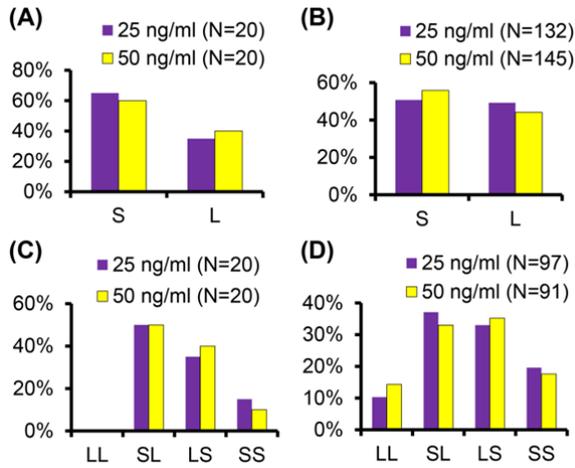

**Figure 4.** (A) Directional choices of the first cells to reach the maze bifurcation (S=short path, L=long path). (B) Directional choices of all cells reaching the maze bifurcation, regardless of order. (C) Directional choices of the first two cells reaching the maze bifurcation, when the cell sequence is taken into account. (D) Directional choices of any two consecutive cells to reach the maze bifurcation.

In order to investigate this curious phenomenon further, we considered the sequence in which the cells arrived at the bifurcation. This type of analysis showed that the cells alternate their directional choices between the long and the short paths (see Videos S1 and S2). In fact, over 90% of trailing cells chose a different path compared to the leading cell, when only the first two cells that entered the maze were considered. Figure 4C shows that there was a slight bias (~50% of the occurrences) towards the short-long decision, because the leading cells were likelier to choose the short path (as was shown in Figure 4A). The second likeliest case was long-short; followed by short-short, accounting for less than 10% of the occurrences. Finally, the long-long case happened with a negligible probability.

Moreover, the directional choice alternation persisted even with the subsequent cells. Figure 4D shows the probabilities of fibroblast directional decisions when the choices of any two consecutive cells are considered. Here, approximately 70% of the following cells made the opposite choice to their predecessor, and only 30% of them followed the cell before. For simplicity, we only examined non-dividing cells when collecting data for Figure 4. However, since tissue formation is a highly proliferative process, it is also interesting to examine how cell mitosis affects the fibroblast chemotaxis.

In the past, mitosis has been found to affect fibroblast migration under non-chemotactic conditions,[12, 13] while here we explored its effect on the cells' directional decision-making in the presence of PDGF-BB. To do that, the fibroblasts undergoing division were manually tracked and analyzed. Figure 5A shows a time montage of such an event occurring, while the mother cell was navigating the maze's common path from the entrance. This figure shows that after the division, one of the daughter cells continued along the positive PDGF-BB gradient path, while the second daughter went against it (in the opposite direction).

Overall, we observed two modes of the daughter cells' migration: either both went along the gradient (though often taking different paths through the maze, as in the cell sequence case), or one went along the gradient and the other against it (see Figure 5B). However, the probabilities of each mode differed significantly: only < 10% of occurrences for the former (Figure 5C, Video S3), versus >90% for the latter (see Video S4). Furthermore, it was especially surprising to see that some of the daughter cells that went against the gradient did so despite the oncoming traffic of other migrating cells; and ultimately were able to exit the maze entirely, through the entrance (see Video S4). Therefore, it is apparent that the mitosis-driven repulsion between the daughter fibroblasts is so great that it is able to overcome both the chemotactic forces and the "inertia" of the other cells migrating along the gradient.

Finally, we observed that the mitosis happened more frequently at regions in the maze that are closer to the entrance, and less frequently near its exit (see Figure S1). This is likely because the proliferation threshold happens at lower levels of PDGF-BB,[22] which correspond to the concentrations experienced by the cells when they enter the maze.

## Discussion

This manuscript investigated fibroblast chemotaxis in tissue-mimicking architectures, in order to understand how these cells influence each other's directional decision-making under microscopic confinement. Understanding these mechanisms is important to developmental biology, multiple tissue pathologies,[36-41] and regenerative medicine,[1] because fibroblasts are the primary cell type responsible for normal tissue homeostatic processes. For example, they are involved in collagen synthesis, the build-up of connective tissues (e.g., cartilage, bone), and wound healing in response to injury.[2, 42] Therefore, characterizing their migratory behavior in microscopic pores can lead to improved tissue patterning and product uniformity. Subsequently, since tissue generation is a highly dynamic and proliferative process, we hypothesized that cell-cell interactions and cell division should affect fibroblasts' decisions.  However, our literature review of fibroblast migration (see **Error! Reference source not found.**) and directional migration (see **Error! Reference source not found.**) studies showed that such processes have not yet been investigated.

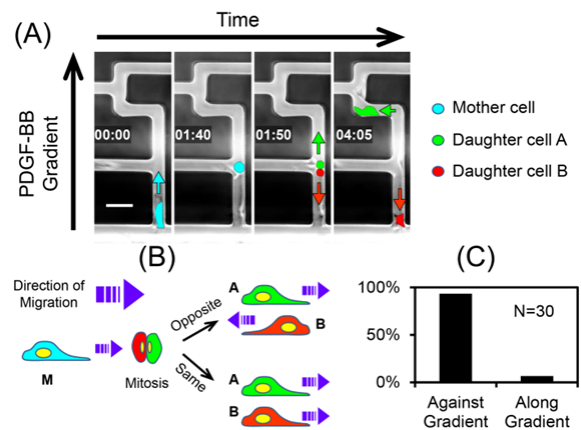

**Figure 5.** (A) Time-lapsed montage of cells undergoing mitosis when migrating in the maze. The cells were false-colored for visual recognition. Cyan, red, and green arrows show the direction of migration of the mother and the two daughter cells, respectively. The time is displayed as hh:mm. The scale bar is 50 μm. (B) Diagram showing two directional scenarios encountered by the two daughter cells. (C) Frequency of occurrence of the two directional scenarios shown in pane B.

To test the hypothesis, we induced fibroblast chemotaxis in microfluidic mazes and collected migration statistics as a function of the sequence at which the cells arrived at the path bifurcations.  Unexpectedly, we found that consecutive cells alternate their decisions in order to avoid following each other, even if that means choosing the longer path through the maze (i.e., a weaker chemoattractant gradient).  This seems to be contradictory to studies performed in similar mazes, but with cancer cells and immune cells.[10, 17] However, a major difference is that those were single cell studies, while ours were multi-cell.  Therefore, one possible explanation is that the leading cell alters the local chemoattractant concentration by consuming the PDGF-BB (at a rate higher than its replenishment via diffusion).  Subsequently, choosing an alternate path may in fact result in a steeper local gradient for the subsequent cells.  Another possible explanation is the integrin migration tracks that fibroblasts have been known to leave behind them.[11, 12]  However, this is less likely, as those findings report cells following each other's migration tracks, not avoiding.  Moreover, fibroblasts are known to attract each other via fibroblast growth factor and ECM degradation proteinase release.[21]  Therefore, it is surprising to see them avoiding each other to such a great extent that they would contradict the optimal route expected from the PDGF-BB gradient.

Additionally, we demonstrated that the division of the parental fibroblasts was highly asymmetric in terms of the migration choices of their descendants. This was evidenced by one of the daughters being more likely to move against the chemoattractant gradient, often disregarding the oncoming traffic created by the other cells in the maze.  The finding is consistent with previous reports of fibroblast daughter cells creating tracks that are mirror images of each other, when migrating on a gold particle-coated substrate;[12] and also, migrating in opposite directions on a collagen matrix.[13] However, in the latter work, the daughters separated only temporarily when the cells divided on fibrin (instead of on collagen), and eventually came back together along the same path (which was not the case in our study).  Therefore, the effect appears to be substrate-dependent. Moreover, these experiments were conducted in the absence of a chemotactic gradient and without micro-confinement, so they are not a one-to-one comparison with our work.

Such heterogeneity between sibling cells is typically attributed to differences in the relative expressions of chemoattractant receptors on their surfaces.  The case of the fibroblast PDGF-BB-driven chemotaxis is also a receptor-mediated process.[21, 43] Therefore, it is possible that there were differences in PDGF receptor expression levels between the sibling cells.  These, in turn, could have resulted from an uneven distribution of the receptors between the anterior and posterior parts of the mother cells' membranes, during the chemotaxis (see Figure 6).  Hence, it would be interesting to image the receptor distribution in the fibroblasts membranes, at the time of mitosis. However, immunofluorescent labelling of the PDGF receptors would also block the receptors sites, subsequently hindering the chemotaxis.

In summary, our study provides insight into how individual fibroblasts affect each other's decision-making processes during chemotaxis in tissue-like microarchitectures.  The results of this work carry practical implications for both understanding the biology of chemotactically-driven morphogenesis processes in vivo, as well as for artificial tissue design in vitro.  For example, scaffold architectures could be optimized for achieving desired cell distributions by taking into account how the fibroblasts alternate the paths they take at bifurcations.  Furthermore, our model could help to select cells with a higher migration potential by selectively isolating the cells that choose the shorter path through the maze.  Finally, this study raises some critical questions about what is behind the discovered effects that compete with the chemotaxis.  For example, the reason why the mitotic cells display highly distinctive directional choices may be either due to a difference in receptor expression in the daughter cells, or due to the migration tracks left by their parent.  Therefore, these effects, and other possibilities, need to be investigated further.

## Conclusions

In conclusion, we conducted a study of fibroblast interactions during chemotaxis in a microfluidic maze made to resemble in vivo tissue pores. Through this study, we have demonstrated that the directional decisions of these cells are influenced by the sequence in which they arrive at path bifurcations, and that their path choices alternate depending on each predecessor's decision.  Also, we showed that cell division occurring during the chemotaxis yields daughter cells with directional bias distinctive from that of their siblings. Furthermore, in both cases, the fibroblast-fibroblast influence appeared to overcome the directional guiding of the chemoattractant gradient established within the maze. We presented a hypothesis the that sequence effect is likely due to a localized chemoattractant consumption by the leading cells, while the mitotic effect could be caused by an asymmetric PDGF-BB receptor inheritance by the two daughters. However, further investigations are needed to validate the hypotheses. Nonetheless, the presented results carry practical implications for developmental biology, multiple pathologies, and tissue engineering.

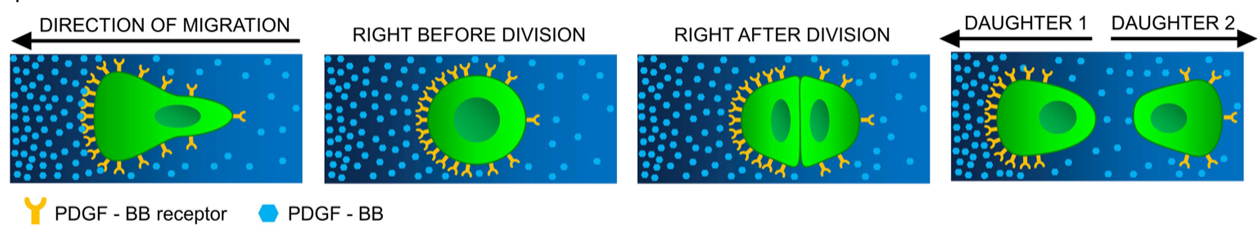

**Figure 6**. Asymmetric distribution of PDGF-BB receptors over the surface of a chemotaxing fibroblast in response to a PDGF-BB gradient leads to an asymmetric division, generating daughter cells of different directional choices and sensing capability.

## Acknowledgements

This study was funded by the Gustavus and Louise Pfeiffer Research Foundation. The authors also thank NJIT's McNair Achievement and Provost Summer Research Programs for providing student labor for this project. Finally, fibroblast donation from Prof. Xiaoyang Xu's laboratory at NJIT's Department of Chemical, Biological and Pharmaceutical Engineering is greatly appreciated.

## References

1. T. Wong, J. A. McGrath and H. Navsaria, *Br J Dermatol*, 2007, **156**, 1149-1155.
2. G. Sriram, P. L. Bigliardi and M. Bigliardi-Qi, *European journal of cell biology*, 2015, **94**, 483-512.
3. S. Kurosaka and A. Kashina, *Birth defects research. Part C, Embryo today : reviews*, 2008, **84**, 102-122.
4. G. Reig, E. Pulgar and M. L. Concha, *Development*, 2014, **141**, 1999-2013.
5. C. M. Franz, G. E. Jones and A. J. Ridley, *Developmental cell*, 2002, **2**, 153-158.
6. E. Scarpa and R. Mayor, *The Journal of cell biology*, 2016, **212**, 143-155.
7. W.-J. Rappel, *Proceedings of the National Academy of Sciences*, 2016, **113**, 1471-1473.
8. D. Ellison, A. Mugler, M. D. Brennan, S. H. Lee, R. J. Huebner, E. R. Shamir, L. A. Woo, J. Kim, P. Amar, I. Nemenman, A. J. Ewald and A. Levchenko, *Proceedings of the National Academy of Sciences of the United States of America*, 2016, **113**, E679-E688.
9. X. Trepat, Z. Chen and K. Jacobson, *Comprehensive Physiology*, 2012, **2**, 2369-2392.
10. C. Scherber, A. J. Aranyosi, B. Kulemann, S. P. Thayer, M. Toner, O. Iliopoulos and D. Irimia, *Integrative biology : quantitative biosciences from nano to macro*, 2012, **4**, 259-269.
11. G. Kirfel, A. Rigort, B. Borm and V. Herzog, *European journal of cell biology*, 2004, **83**, 717-724.
12. G. Albrecht-Buehler, *Cell*, 1977, **11**, 395-404.
13. M. Miron-Mendoza, X. Lin, L. Ma, P. Ririe and W. M. Petroll, *Experimental eye research*, 2012, **99**, 36-44.
14. M. Mak and D. Erickson, *Lab on a chip*, 2014, **14**, 964-971.
15. G. Costa, K. I. Harrington, H. E. Lovegrove, D. J. Page, S. Chakravartula, K. Bentley and S. P. Herbert, *Nature cell biology*, 2016, **18**, 1292-1301.
16. J. Yan and D. Irimia, *Technology*, 2014, **2**, 185-188.
17. V. Ambravaneswaran, I. Y. Wong, A. J. Aranyosi, M. Toner and D. Irimia, *Integrative biology : quantitative biosciences from nano to macro*, 2010, **2**, 639-647.
18. H. Seppä, G. Grotendorst, S. Seppä, E. Schiffmann and G. R. Martin, *The Journal of cell biology*, 1982, **92**, 584-588.
19. D. I. Shreiber, P. A. Enever and R. T. Tranquillo, *Experimental cell research*, 2001, **266**, 155-166.
20. A. Siegbahn, A. Hammacher, B. Westermark and C. H. Heldin, *The Journal of clinical investigation*, 1990, **85**, 916-920.
21. A. Albini, B. C. Adelmann-Grill and P. K. Muller, *Collagen and related research*, 1985, **5**, 283-296.
22. A. De Donatis, G. Comito, F. Buricchi, M. C. Vinci, A. Parenti, A. Caselli, G. Camici, G. Manao, G. Ramponi and P. Cirri, *The Journal of biological chemistry*, 2008, **283**, 19948-19956.
23. J. Lepisto, M. Laato, J. Niinikoski, C. Lundberg, B. Gerdin and C. H. Heldin, *The Journal of surgical research*, 1992, **53**, 596-601.
24. J. Lepisto, J. Peltonen, M. Vaha-Kreula, J. Niinikoski and M. Laato, *Biochem Biophys Res Commun*, 1995, **209**, 393-399.
25. S. Vedel, S. Tay, D. M. Johnston, H. Bruus and S. R. Quake, *Proceedings of the National Academy of Sciences*, 2013, **110**, 129-134.
26. M. C. Leong, S. R. K. Vedula, C. T. Lim and B. Ladoux, *Communicative & Integrative Biology*, 2013, **6**, e23197.
27. M. Abercrombie and J. E. Heaysman, *Experimental cell research*, 1953, **5**, 111-131.
28. M. Abercrombie and J. E. Heaysman, *Experimental cell research*, 1954, **6**, 293-306.
29. B. da Rocha-Azevedo and F. Grinnell, *Experimental cell research*, 2013, **319**, 2440-2446.
30. D. Levinstone, M. Eden and E. Bell, *Journal of cell science*, 1983, **59**, 105-119.
31. H. V. Prentice-Mott, C. H. Chang, L. Mahadevan, T. J. Mitchison, D. Irimia and J. V. Shah, *Proceedings of the National Academy of Sciences of the United States of America*, 2013, **110**, 21006-21011.
32. C. D. Paul, D. J. Shea, M. R. Mahoney, A. Chai, V. Laney, W. C. Hung and K. Konstantopoulos, *FASEB journal : official publication of the Federation of American Societies for Experimental Biology*, 2016, **30**, 2161-2170.
33. S. Rao, U. Tata, V. Lin and J.-C. Chiao, *Micromachines*, 2014, **5**, 13.
34. C. A. Schneider, W. S. Rasband and K. W. Eliceiri, *Nature methods*, 2012, **9**, 671-675.
35. B. Akar, B. Jiang, S. I. Somo, A. A. Appel, J. C. Larson, K. M. Tichauer and E. M. Brey, *Biomaterials*, 2015, **72**, 61-73.
36. T. Kisseleva and D. A. Brenner, *Exp Biol Med (Maywood)*, 2008, **233**, 109-122.
37. T. Kisseleva and D. A. Brenner, *Proceedings of the American Thoracic Society*, 2008, **5**, 338-342.

38. O. Z. Lerman, R. D. Galiano, M. Armour, J. P. Levine and G. C. Gurtner, *The American journal of pathology*, 2003, **162**, 303-312.
39. D. A. Beacham and E. Cukierman, *Seminars in cancer biology*, 2005, **15**, 329-341.
40. R. Castello-Cros and E. Cukierman, *Methods Mol Biol*, 2009, **522**, 275-305.
41. A. D. Rouillard and J. W. Holmes, *The Journal of physiology*, 2012, **590**, 4585-4602.
42. S. Rhee, *Experimental & molecular medicine*, 2009, **41**, 858-865.
43. S. T. Christensen, I. R. Veland, A. Schwab, M. Cammer and P. Satir, *Methods in enzymology*, 2013, **525**, 45-58.

# SUPPLEMENTAL MATERIALS

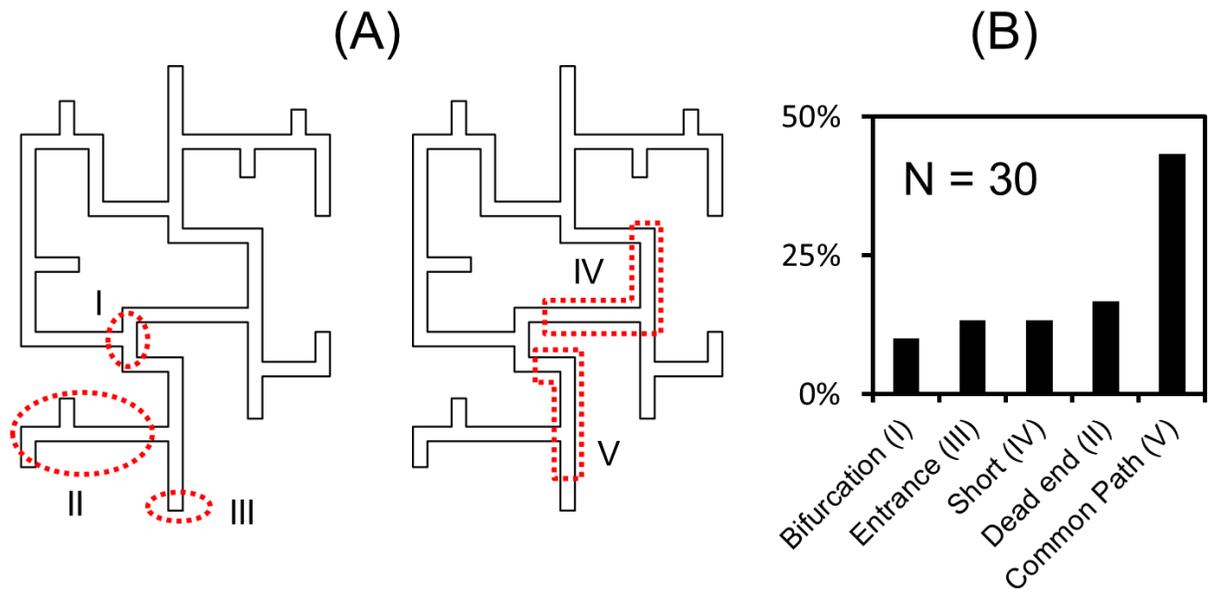

Figure S1. (A) Spatial distribution of mitosis occurrence at different locations in the maze. (B) Frequency of mitosis occurrence at various locations described in A. N indicates the number of cells being tracked.

Video S1. First cells selecting short path and alternative pattern of cell decision making

https://vimeo.com/234413123

Video S2. First cells selecting long path and alternative pattern of cell decision making

https://vimeo.com/234413129

Video S3. Cell division at leading channels and follow each other

https://vimeo.com/234413134

Video S4. Cell division at the leading channel

https://vimeo.com/234413136